\documentstyle[twoside,fleqn,espcrc2,epsf]{article}

\newcommand{\hs}{\hspace{0.1cm}}

\title{Progress using generalized lattice Dirac operators to parametrize the Fixed-Point QCD action}
\author{P.\ Hasenfratz\address[BERN]{Institute for Theoretical Physics,
    University of Bern, Sidlerstrasse 5, CH-3012 Bern,
    Switzerland}\thanks{Based on a talk by K.H. and a poster by T.J.}, 
    S.\ Hauswirth\addressmark, K.\ Holland\addressmark, 
    T.\ J\"org\addressmark, F.\ Niedermayer\addressmark, 
    U.\ Wenger\address[OXFORD]{Theoretical Physics, University of Oxford, 1
    Keble Road, Oxford, OX1 3NP, England}}

\begin{document}

\begin{abstract}
We report on an ongoing project to parametrize the Fixed-Point Dirac 
operator for massless quarks, using a very general construction which has arbitrarily 
many fermion offsets and gauge paths, the complete Clifford algebra and
satisfies all required symmetries. Optimizing a specific construction 
with hypercubic fermion offsets, we present some preliminary
results. 
\vspace{1pc}
\end{abstract}

\maketitle

\section{Motivation for improved fermionic actions}

There are many reasons to seek an improved lattice Dirac operator. The most
commonly used one, the Wilson Dirac operator, has a number of problems. Chiral
symmetry, an essential element of QCD, is broken explicitly by a term which 
removes the fermion-doublers and so the pion is not massless for zero quark 
mass. The pion can be tuned massless by varying the hopping
parameter. However, this does not remove other problems, such as the
mixing of operators in different chiral representations, which makes the
extraction of chiral matrix elements difficult. There are also problems
measuring chiral quantities such as the quark condensate and zero modes of the
Dirac operator and connecting these to the topology of the gauge fields, due
to, among other things, topological defects and exceptional configurations. In
addition, the Wilson Dirac operator has large lattice artifacts. 

A solution to problems of chirality has been proposed by Ginsparg and Wilson
\cite{Gin82}: a lattice Dirac operator which satisfies
\begin{equation}
\{ \gamma_5, D^{-1} \} = 2 \hs a R \hs \gamma_5, 
\label{eq:GW} 
\end{equation}
where $R$ is any local function which commutes with $\gamma_5$ and $a$ is the
lattice spacing, was expected to define a lattice theory with the same chiral
properties as the continuum theory. Unfortunately, no solution to this
relation was found and the idea was abandoned. The first lattice
regularization of fermions with chiral symmetry, the Domain Wall fermions
\cite{Kap92} and the related Overlap construction \cite{Nar93}, followed a different path
and seemed to be unrelated to the Ginsparg-Wilson (GW) relation.

Following the observation that the Fixed-Point Dirac operator satisfies the GW
relation \cite{Has98}, the interest turned to this general formulation again. The GW
relation turned out to be a powerful theoretical tool: it implies an index
theorem on the lattice \cite{Lal98} and all the chiral properties of the formal
continuum theory \cite{Has98b}. This last feature became trivial after it was
recognized that an exact chiral symmetry transformation follows from
eq.(\ref{eq:GW}) \cite{Lus98}. The observation that the Overlap Dirac operator
satisfies the GW relation connected the Domain Wall approach to this general
formulation \cite{Neu98}. The progress in vector chiral symmetry was followed by a
breakthrough in the formulation of regularized chiral gauge theories also
\cite{Lus99}. 

All these developments have led to much recent activity on the theoretical
understanding of lattice chiral symmetry \cite{Nie99,Cha99} and on constructing improved
lattice Dirac operators which approximately satisfy the GW relation \cite{Deg00,Gatt00,Haus00}. At
the same time, detailed tests and the first large-scale simulations with the
simplest version of Domain Wall and Overlap Dirac operators demonstrated the
power and difficulties of working with chirally symmetric QCD actions \cite{Neu98b}.

\section{General lattice Dirac operators}

Before we explain our approach to the parametrization of the Fixed-Point Dirac operator
we briefly discuss the structure of general Dirac operators which satisfy the
basic symmetries on the lattice. This will make clear what kind of lattice 
operators can be used in the parametrization of the Fixed-Point Dirac operator or in any 
other parametrization of a Dirac operator. A more detailed discussion of these
issues is given in \cite{Haus00}. A general gauge covariant lattice operator with
color, space and Dirac indices can be written as:   
\begin{equation}
  \label{Dgen}
  D = \sum_A \Gamma_A \sum_l c(\Gamma_A,l) \hat{U}(l) \, ,
\end{equation}
where the $\Gamma_A$ are elements of the Clifford algebra basis which we define as
$\Gamma = 1, \gamma_\mu, i \sigma_{\mu \nu}, \gamma_5, \gamma_\mu \gamma_5$,
$l$ denotes a path,
$\hat{U}(l)$ the product of the link matrices along this path and
$c(\Gamma_A,l)$ is the coupling
for the given path and Clifford algebra element. 

The basic symmetries of the Dirac operator, which are gauge symmetry,
$\gamma_5$-hermiticity, charge conjugation, hypercubic rotations and
reflections, impose the following restrictions on 
eq.(\ref{Dgen}). 

\begin{itemize}
\item Translation invariance requires that the Dirac operator $D(n,n+r)$
  connecting lattice sites $n$ and $n+r$ depends on $n$ only through the
  $n$-dependence of the gauge field, i.e.\ the couplings $ c(\Gamma_A,l)$ 
  are constants or gauge invariant functions of gauge fields, respecting 
  locality and invariance under the symmetry transformations.
  
\item $\gamma_5$-hermiticity and charge conjugation together imply that the couplings $ c(\Gamma_A,l)$ 
  for our choice of the Clifford algebra basis are real.
  Further, from hermiticity it follows that the path $l$ and the opposite path $\bar{l}$ 
  (or equivalently,  $\hat{U}(l)$ and  $\hat{U}(l)^\dagger$) 
  should enter in the combination
  \begin{equation}
    \label{GUUd}
    \Gamma \left( \hat{U}(l)+\epsilon_\Gamma \hat{U}(l)^\dagger \right) \,,
  \end{equation}
  where the sign $\epsilon_\Gamma$ is defined by
  $\gamma^5 \Gamma^\dagger \gamma^5 =\epsilon_\Gamma \Gamma$.
  
\item Permutations and reflections of the coordinate axes (hypercubic rotations) imply 
  that for a given path $l_0$ a whole class of paths belongs to the Dirac
  operator. These paths are related to $l_0$ by all the $16 \times 24 = 384$ reflections and permutations
  of the coordinate axes. Under such a symmetry transformation $\alpha=1,\ldots,384$ the Clifford algebra 
  element $\Gamma_0$ associated with $l_0$ generally transforms into $\Gamma^{(\alpha)}$. Furthermore the sign 
  of the couplings may change, whereas their absolute value remains unchanged.
\end{itemize}

A Dirac operator satisfying all the basic symmetries can be written as
\begin{eqnarray}
\label{sym}
D &=& \sum_{\Gamma_0,l_0} c(\Gamma_0,l_0) \sum_\alpha O^\alpha \nonumber \\
O^\alpha &=& \Gamma^{(\alpha)} \left( \hat{U}(l^{(\alpha)}) +
\epsilon_\Gamma \hat{U}(l^{(\alpha)})^\dagger \right) \,.
\end{eqnarray}
where the sum runs over a set of reference paths defined by $\Gamma_0$ and $l_0$ as well as over all
the symmetry transformations $\alpha$ defined by the group of permutations and reflections of the coordinate axes.

It is quite natural to include the full Clifford algebra in the Dirac
operator. The Wilson operator already contains $1$ and $\gamma_\mu$, the
$O(a)$ Symanzik condition involves $\sigma_{\mu \nu}$ and the topological
charge is proportional to $\mbox{Tr} (\gamma_5 D)$, which is zero unless $D$
contains $\gamma_5$. Furthermore, for $\gamma_5$-hermitian Dirac operators, 
the Ginsparg-Wilson relation (\ref{eq:GW}) can be written as 
\begin{equation}
\label{eq:GW2}
D + D^{\dagger} = 2 \hs a \hs D^{\dagger} R D,
\end{equation}
which can only be satisfied if $D$ contains the complete Clifford algebra.

\section{Efficient implementation of general Dirac operators}

On first appearances one might think that it is not feasible to calculate 
such a general Dirac operator
which has many different couplings, where every coupling might contain as 
many as $768$ paths. But one has to
keep in mind that the calculation of propagators for small quark masses needs 
several hundreds or even thousands 
of conjugate gradient steps and therefore one can afford to spend some time to
precalculate and store the whole Dirac
operator before one starts the calculation of the propagator. On top of this 
there are two reasons why the building 
of general operators can be done in a very efficient way:

\begin{itemize}

\item There are a lot of paths which are invariant under certain subgroups of the reflections and 
  permutations which reduces the number of terms significantly and in some case even
  leads to a cancellation of certain terms because they have relative minus signs. 
\item A less trivial fact is that the sum of paths for many couplings 
  can be factorized in an efficient way, i.e.\ that such large sums of many
  paths can be written as a product of smaller sums of fewer paths.
\end{itemize}

As an example we consider a nearest neighbour coupling with $\Gamma_0 =
\gamma_5$ and generating path $l_0 = [2,1,-2,3,4,-3,-4]$, where a path 
$l$ is  a sequence of steps in the $\pm \mu$ direction. All the paths of 
this coupling can be written in the following compact way:
\begin{equation}
  \label{g5_1}
  \gamma_5 \sum_{\mu\nu\rho\sigma} \epsilon_{\mu\nu\rho\sigma}
  \left( S_{\mu \nu} P_{\rho\sigma}
    + P_{\rho\sigma} S_{\mu \nu} + \mbox{h.c.} \right) \,,
\end{equation}
where the color matrices $S_{\mu \nu}$ and $P_{\rho\sigma}$ are
certain combinations of staples or plaquettes, respectively. 
When all the plaquettes and staples and the most frequent combinations, e.g.\ $P_{\rho\sigma}$, 
are precalculated then operators as in eq.(\ref{g5_1}) can be calculated very quickly. As an illustration of this we 
consider one of our parametrizations of the Fixed-Point Dirac operator. It has 39 couplings in total,
at least one per offset on the 1-hypercube and per type of Clifford algebra
element. Building this operator takes roughly 15 times as long as to multiply the operator with a vector
and therefore it is a very small fraction of the time used to perform a calculation of one propagator.

The basic operation required to calculate propagators or eigenvalues is the
multiplication of the Dirac operator times a vector. For a Dirac
operator with hypercubic fermion offsets which contains the complete Clifford
algebra, the matrix times vector multiplication is roughly 40 times more expensive
than for the Wilson Dirac operator.   

In \cite{Haus00}, we have included a {\tt MAPLE} code which, given an input path $l$
and any Clifford algebra element $\Gamma$, gives all of the paths
$l^{(\alpha)}$ and algebra elements $\Gamma^{(\alpha)}$ generated by all
permutations and reflections of the coordinate axes. This can be used by
anyone who wishes to use a very general lattice Dirac operator, independently
of how this operator is optimized. 

\section{Fixed-Point improved actions}

\begin{figure}[tb]
\epsfxsize=\hsize
\epsfbox{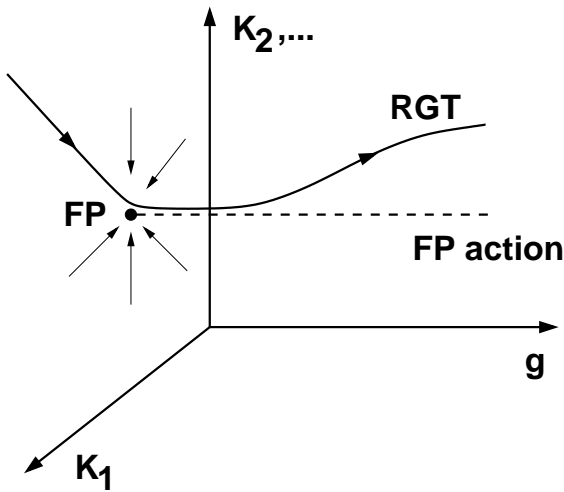}
\caption{Renormalization Group trajectory of asymptotically-free theories.}
\label{fig:rgt}
\end{figure}

The Fixed-Point method of improving lattice actions is inspired by the
Renormalization Group flow of asymptotically-free theories \cite{Has94}. Consider a lattice
action which contains all possible interactions. The Renormalization Group is
defined by some blocking function which averages over the fields to produce a
new action on a coarser lattice with fewer fields. The new action typically
has different couplings from the original action and so we can imagine the
blocking step as a flow in the coupling parameter space. Infinitely many
blocking steps give the complete Renormalization Group (RG) trajectory. In
Figure \ref{fig:rgt}, we show the RG trajectory for QCD with massless
quarks. The Fixed-Point (FP) has the property that the couplings are
reproduced after a blocking step. For asymptotically-free theories, the
Fixed-Point is on the surface of vanishing coupling $g = 0$. 

If one starts on this surface, the RG trajectory flows quickly to the
Fixed-Point. If one starts close to this surface at some small coupling $g$
(i.e. small lattice spacing $a$), the RG trajectory flows quickly towards the 
Fixed-Point and then flows away from it. Starting on the RG trajectory with
arbitrarily fine lattices with arbitrarily small lattice artifacts, one can
reach any point on the trajectory by making sufficiently many blocking
steps. All actions on the trajectory describe the same physics. The
physical observables of the continuum quantum theory are identical to those of
any lattice quantum theory on the RG trajectory, independently of the lattice
coarseness. Such lattice actions are said to be quantum perfect. The
Fixed-Point action is an approximation to the RG trajectory for small
couplings $g$ and is classically perfect, i.e. it completely describes the
continuum classical theory without discretization errors. 

Fixed-Point actions have many desirable features. By closely approximating the
RG trajectory, they are expected to have much reduced quantum lattice
artifacts. They can be optimized for locality. The Fixed-Point Dirac operator
satisfies the Ginsparg-Wilson relation and so has good chiral behavior. The
Fixed-Point QCD action has well-defined topology and satisfies the index
theorem on the lattice. Although more expensive than standard actions,
studies of other models have shown that practical Fixed-Point actions 
can be constructed and used \cite{Bla95}. 

\section{Free Fixed-Point fermions}

\begin{figure}[tb]
\epsfxsize=\hsize
\epsfbox{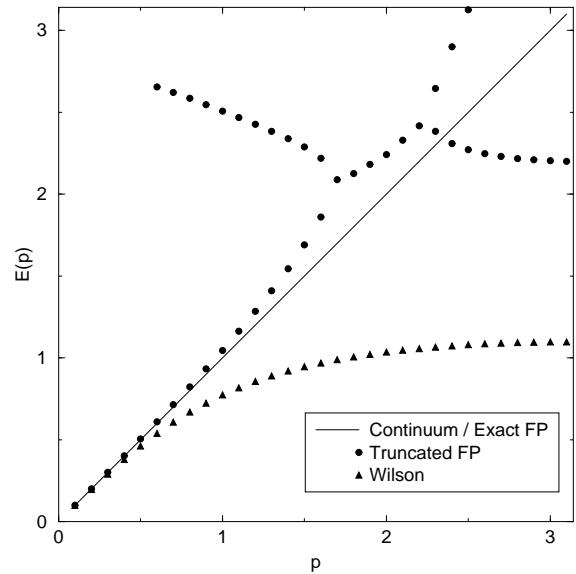}
\caption{Free fermion dispersion relation.}
\label{fig:dispersion}
\end{figure}

We first consider free massless fermions.
Because the fermionic action is quadratic in the fermionic fields, the
Renormalization Group step for the fermionic fields amounts to Gaussian
integration, which can be done exactly. The blocking step connects the 
Dirac operator on the fine (F) and coarse (C) lattices by
\begin{equation}
  \label{RGeq}
  D^{-1}_{\rm C} = \frac{1}{\kappa} + \Omega \hs D^{-1}_{\rm F} \hs \Omega^{\dagger},
\end{equation}
provided $D_{\rm F}$ has no zero modes, where $\kappa$ is an optimizable free
parameter of the blocking and $\Omega$
is the blocking function used to integrate out the fine fields to produce the 
coarse fields. The FP Dirac operator is reproduced under the blocking step
i.e. $D_{\rm C} = D_{\rm F} = D_{\rm FP}$. For free fermions, this equation can be solved exactly
analytically. The FP Dirac operator is local, i.e. the couplings
fall off exponentially with distance, and the rate of fall off can be maximized by
varying $\kappa$. However, $D_{\rm FP}$ contains infinitely
many couplings. For practicality, $D_{\rm FP}$ is approximated with an ultra-local
operator, for which each point is only coupled to its neighbours on the
hypercube. In Figure \ref{fig:dispersion}, we see the exact $D_{\rm FP}$ 
reproduces exactly the continuum dispersion relation, while the approximate 
FP Dirac operator has much smaller discretization errors than the Wilson
operator. The unphysical branches in the approximate $D_{\rm FP}$ occur at large 
momenta and only have a small effect.

\begin{figure}[tb]
\epsfxsize=\hsize
\epsfbox{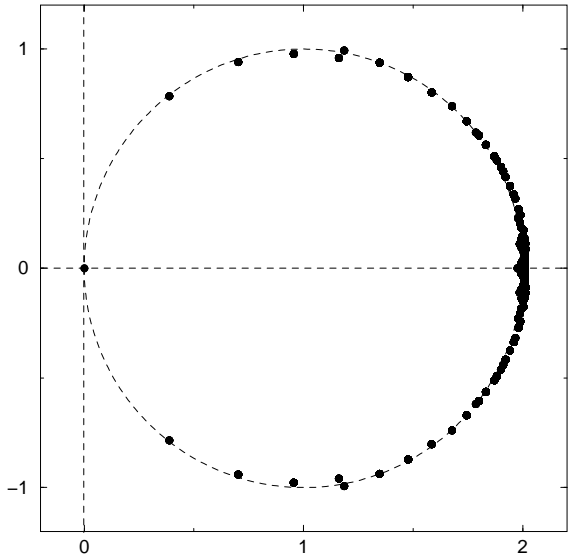}
\caption{Free fermion eigenvalue spectrum of the truncated FP Dirac operator.}
\label{fig:spectrum}
\end{figure}

The Ginsparg-Wilson relation eq.(\ref{eq:GW}) is a constraint on $\{ \gamma_5,
D^{-1} \}$. The FP equation (\ref{RGeq}) connects the propagators
$D^{-1}$ on the coarse and fine lattices. Combining these equations gives
\begin{equation}
  \label{RGeqR}
  R_{\rm C} = \frac{1}{\kappa} + \Omega \hs R_{\rm F} \hs \Omega^{\dagger},
\end{equation}
and at the Fixed-Point, $R_{\rm C} = R_{\rm F} = R_{\rm FP}$. For free fermions, this equation
can also be solved exactly analytically. Choosing a symmetric overlapping
block transformation $\Omega$ with a scale factor 2 averaging over 
hypercubes \cite{Kun98}, the exact $R_{\rm FP}$ is ultra-local and has only
hypercubic couplings --- no approximation for practicality is required. The 
block transformation $\Omega$ determines $R$. With other methods to build a 
Dirac operator satisfying the Ginsparg-Wilson relation, for example the
Overlap construction, $R$ is unconstrained and typically $R = \frac{1}{2}$. 

Defining a rescaled Dirac operator $d = \sqrt{2 R} \hs D \hs \sqrt{2 R}$, the
Ginsparg-Wilson relation can be written (with lattice spacing $a = 1$)
\begin{equation}
  \label{circle}
  d + d^{\dagger} = d^{\dagger} \hs d,
\end{equation}
i.e. the eigenvalues of $d$ lie on a circle of radius 1 centred at
$(1,0)$. Using $R_{\rm FP}$ and the hypercubic approximation of $D_{\rm FP}$, we show
the eigenvalue spectrum of $d$ in Figure \ref{fig:spectrum}. It lies almostly 
exactly on the circle, indicating that the hypercubic truncation has only 
slightly affected the Ginsparg-Wilson property.

\section{Parametrization of the QCD FP Dirac operator}

To parametrize the QCD FP Dirac operator $D_{\rm FP}$ for massless quarks, 
we use a general Dirac operator as defined in eq.(\ref{sym}) with all the
couplings of the $1$-hypercube and all elements of the Clifford algebra. 
The gauge invariant functions we use 
as couplings of $D_{\rm FP}$ are polynomials in local fluctuations of the gauge fields. 
Furthermore we include the possibility to smear the gauge fields and to 
project them back to SU(3), i.e.\ we are 
using so called fat links. We use the new parametrization of the FP gauge
action \cite{Ruf00} which also makes use of fat gauge links.

The QCD FP action is also quadratic in the fermion fields and the Renormalization Group
step can again be done analytically. The QCD FP equation is the generalization
of the free equation (\ref{RGeq}) given by including gauge fields. In case
$D_{\rm F}$ has zero modes, the QCD FP equation is most conveniently written as 
\begin{eqnarray}
\label{RGO}
D_{\rm C}(V) &=& \kappa \, {\bf 1} \\
&& \hspace{-1.5cm} - \hspace{0.2cm} \kappa^2 \, \Omega(U) [ D_{\rm F}(U) + \kappa \,
\Omega^\dagger(U)\Omega(U) ]^{-1} \Omega^\dagger(U) \, , \nonumber
\end{eqnarray}
where $\kappa$ is an optimizable free parameter of the block transformation
and $U$ and $V$ are the gauge fields on the fine and 
coarse lattice, respectively. They are related through the FP equation of 
the pure SU(3) gauge theory
\begin{equation}
  S^{\rm FP}(V)=\min_{ \{U\} } \left( S^{\rm FP}(U) +T(U,V)\right),
  \label{RGG}
\end{equation}
where $S^{\rm FP}$ is the FP action of the pure SU(3) gauge theory and $T(U,V)$ is the blocking kernel of the block
transformation. As the fluctuations of the gauge fields on the fine lattice are much smaller than those of the coarse gauge
fields, the Dirac operator used on the fine lattice has to have good chiral properties on gauge fields with small 
fluctuations. This however is by far easier than to have a Dirac operator with good chiral properties on gauge 
fields with large fluctuations. An important fact for the parametrization of the FP Dirac operator is that
eq.(\ref{RGO}) can also be given in terms of the propagators
\begin{equation}
  D_{\rm C}^{-1}(V) = \frac{\bf 1}{\kappa} + \Omega(U) D_{\rm F}^{-1}(U) \Omega^\dagger(U),
  \label{RGP}
\end{equation}
as long as $D_{\rm F}$ has no zero modes. In contrast to eq.(\ref{RGO}) the equation for the propagator gives much more 
weight to the small physical modes of the Dirac operator and can therefore be used to improve the properties of
the small modes of the parametrized FP Dirac operator.

The parametrization is an iterative procedure.
We first start at a large value of $\beta$, generate thermal coarse gauge
configurations $V$ with the FP gauge action and determine the corresponding
fine configurations $U$ via minimization as in eq.(\ref{RGG}). Using the free FP
Dirac operator on the fine configurations (which have very small
fluctuations), $D_{\rm C}(V)$ and $D_{\rm C}^{-1}(V)$ are calculated from eqs.(\ref{RGO}) and
(\ref{RGP}). The couplings of the parametrized Dirac operator $D_{\rm par}$ are
determined by minimizing the following $\chi^2$-function:
\begin{eqnarray}
  \label{eq:chi_function}
  \chi^2 &=& \sum || D_{\rm par} v - D_{\rm C} v ||^2 + \nonumber \\
&& \lambda \sum || D^{-1}_{\rm par} v - D^{-1}_{\rm C} v ||^2 \, ,
\end{eqnarray}
where the sum runs over a number of random vectors $v$ on different configurations and $\lambda$ is  
a weighting factor. The use of vectors for the calculation of a $\chi^2$
function for $D_{\rm par}$ is mandatory because the 
definition of $D_{\rm C}$ requires a matrix inversion which we can afford only for a limited number of vectors. 
The minimization of the $\chi^2$ yields a parametrized FP Dirac operator
$D_{\rm par}(V)$ which has good chiral properties over a 
larger range of gauge couplings than the initial truncated free FP Dirac
operator. 

The fitted parametrized operator $D_{\rm par}$ is now used on fine configurations $U'$, determined via
minimization from coarse configurations $V'$ generated thermally at a smaller
value of $\beta$. Minimizing the $\chi^2$ in eq.(\ref{eq:chi_function}) again gives
$D_{\rm par}(V')$ which performs well on an even larger range of gauge couplings. The
whole procedure is repeated until we reach $\beta = 3.0$ which corresponds to 
$\beta_{\rm Wilson} \approx 5.75$.
During this procedure we keep leading terms in the naive continuum limit fixed such that the tree level mass is zero,
the $O(a)$ Symanzik condition is fulfilled, the dispersion relation starts with slope $1$ and the normalization of
the topological charge is correct \cite{Haus00}. Furthermore we fix the free field limit such that we recover the truncated free 
FP operator on the trivial gauge configuration.

In order to parametrize the operator $R$ in eq.(\ref{RGeqR}) we proceed in a similar way as for the Dirac operator. We also 
use a general operator with fat links and fluctuation polynomials. The parametrization of $R$ is however simpler as 
it is trivial in Dirac space and therefore contains a smaller number of operators. In contrast to the equation for the
block transformation of $D_{\rm FP}$  the corresponding equation for $R$ eq.(\ref{RGeqR}) contains no inversion and therefore
the $\chi^2$-function which we minimize in order to find the optimal
parametrization of $R_{\rm par}$ can be defined as
\begin{equation}
  \label{eq:chi_function_r}
  \chi^2 = || R_{\rm par} - R_{\rm C} ||^2 \, ,
\end{equation}
where the norm here is the matrix norm $|| A ||^2 = \sum_{i,j} || a_{i j} ||^2$.

\section{Preliminary results}

\begin{figure}[tb]
\epsfxsize=\hsize
\epsfbox{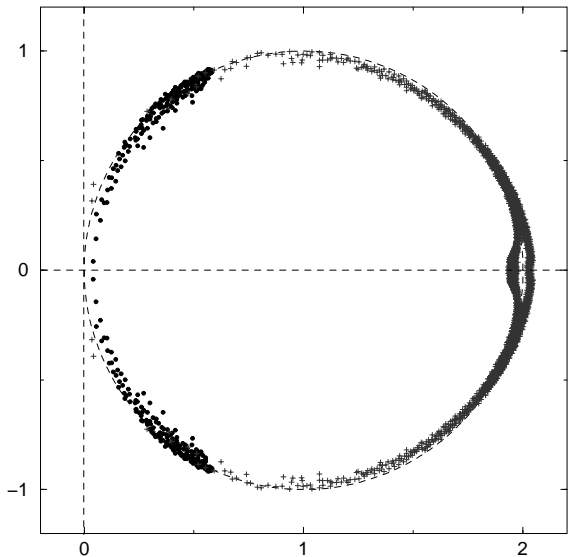}
\caption{Exact eigenvalue spectrum of the parametrized FP Dirac operator at
  $\beta=3.0 \hspace{0.2cm} (\beta_{\rm Wilson} \approx 5.75)$.}
\label{fig:exact_ev}
\end{figure}

We have some preliminary results for the parametrized QCD FP
Dirac operator. In Figure \ref{fig:exact_ev}, we show the spectrum of exact
eigenvalues of $d = \sqrt{2 R} \hs D \hs \sqrt{2 R}$ at $\beta=3.0$, i.e. 
$\beta_{\rm Wilson} \approx 5.75$, corresponding to a lattice spacing of 
$a \approx 0.16 \hspace{0.1cm} {\rm fm}$. We show all the eigenvalues on a
$4^4$ volume (crosses) and those closest to the origin on an $8^4$
volume (circles). We see that the spectrum lies quite close to the Ginsparg-Wilson
circle, with a small additive mass renormalization. This is evidence that the 
parametrization is sufficiently rich to reproduce well the chiral Fixed-Point properties.

In order to measure hadron masses, one needs to calculate the quark propagators.
The quark propagators are determined by solving the equation $X = D^{-1} \eta$ for
some source vector $\eta$. We have measured the accuracy $\epsilon_k$ in calculating
the inverse, $\epsilon_k = ||D(D^{-1} \eta)_k - \eta||$, after $k$ conjugate gradient steps,
i.e. how many times $\eta$ is multiplied by $D$ in calculating $D^{-1}
\eta$. To achieve a given accuracy, we find
that the parametrized FP Dirac operator requires roughly 3 times fewer steps 
than the unpreconditioned Wilson operator corresponding to the same effective 
pion mass for the same gauge configuration. This reduces the overhead of 
$D_{\rm par}$ in production runs.

The chiral properties of a Dirac operator are improved by using the Overlap
construction. Given some input Dirac operator $D_0$, a Dirac operator which 
satisfies the Ginsparg-Wilson relation is given by 
\begin{equation}
\label{overlap}
D_{\rm OV} = 1 - \frac{A}{\sqrt{A^{\dagger} A}}, \hspace{0.2cm} A = 1 - D_0.
\end{equation}
If the input operator already satisfies the Ginsparg-Wilson relation, then
$A^{\dagger} A = 1$ and $D_{\rm OV} = D_0$. The inverse square root is calculated
by some expansion in $A^{\dagger} A$. How quickly the expansion converges is 
determined by the smallest and largest eigenvalues of $A^{\dagger} A$, which
tells us how far the input operator is from satisfying the Ginsparg-Wilson 
relation. 

\begin{figure}[tb]
\epsfxsize=\hsize
\epsfbox{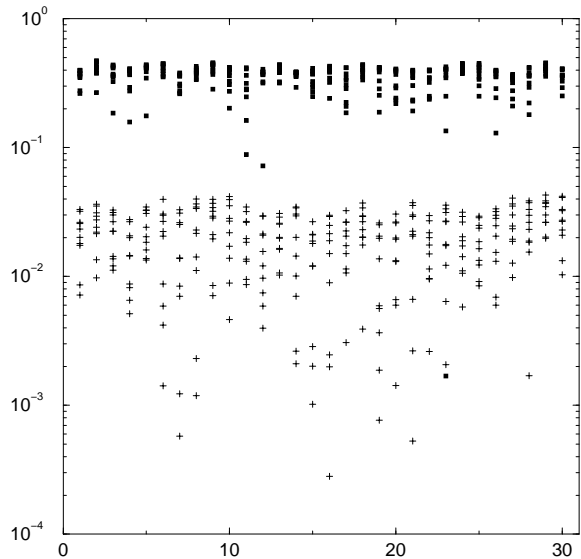}
\caption{10 smallest eigenvalues of $A^{\dagger} A$ at $\beta=3.4 
\hspace{0.2cm} (\beta_{\rm Wilson} \approx 5.95)$ for the parametrized FP (squares) and
Wilson (crosses) Dirac operators for 30 configurations.}
\label{fig:smallest_ev}
\end{figure}

In Figure \ref{fig:smallest_ev}, we plot the 10 smallest eigenvalues of $A^{\dagger} A$ for 
30 configurations of size $8^4$ at $\beta=3.4$, i.e. $\beta_{\rm Wilson} \approx 5.95$,
using $d$ (squares) and $D_{\rm Wilson}$ (crosses) as input. The largest eigenvalues of $A^{\dagger}
A$ are $\sim 1.5$ and $\sim 41$ for $d$ and $D_{\rm Wilson}$ respectively. The
parametrized FP operator $d$ approximately satisfies the Ginsparg-Wilson
relation and the inverse square root can be determined accurately by, for
example, a Legendre expansion with only a few terms, which would very much
improve the already good chiral behavior of $d$. The expansion converges
faster by projecting out the smallest eigenvalues, which are treated
exactly. Using $d$, we see $A^{\dagger} A$ has few small eigenvalues and so
the projection can be done quite cheaply. In contrast, using $D_{\rm Wilson}$ to
construct $A$, most of the eigenvalues of $A^{\dagger} A$ are far from $1$ and
very expensive expansions with up to hundreds of terms are required to
determine the inverse square root \cite{Her00}. There are also more very
small eigenvalues to be projected out to improve the convergence, which would
also be quite expensive.

\begin{figure}[tb]
\epsfxsize=\hsize
\epsfbox{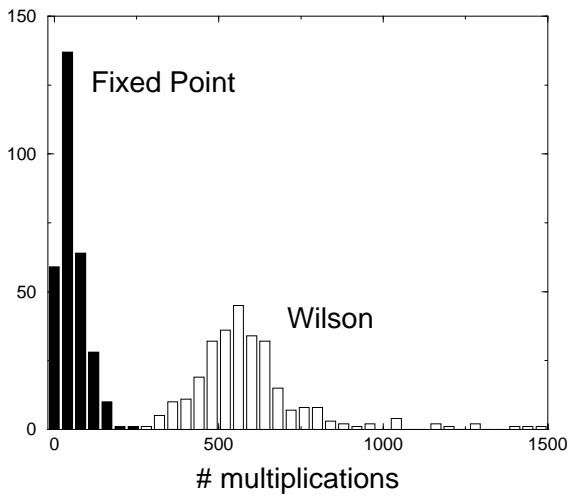}
\caption{Histogram of number of multiplications to find 
10 smallest eigenvalues of $A^{\dagger} A$ for all configurations 
plotted in Figure \ref{fig:smallest_ev}.}
\label{fig:histogram}
\end{figure}

In Figure \ref{fig:histogram}, we plot a histogram of the number
of $A^{\dagger} A$ multiplications required to find the 10 smallest eigenvalues of
$A^{\dagger} A$ using $d$ and $D_{\rm Wilson}$ as input which were plotted in
Figure \ref{fig:smallest_ev}. Using $d$ in the
Overlap construction, fewer small eigenvalues need be treated exactly and
these can be determined with fewer multiplications than are necessary using 
$D_{\rm Wilson}$. 

\section{Conclusions}

The project of parametrizing the QCD FP Dirac operator is ongoing. Preliminary
results show that the parametrization achieved so far gives a Dirac operator
which satisfies the Ginsparg-Wilson relation quite well with a small additive
mass renormalization. Although a matrix vector multiplication with $D_{\rm par}$
is roughly 40 times more expensive than with $D_{\rm Wilson}$, we see that
calculating eigenvalues and propagators converges much faster. The chiral
properties of a Dirac operator are greatly improved by using it as input for the
Overlap construction. As $D_{\rm par}$ is already quite close to satisfying the
Ginsparg-Wilson relation, the Overlap construction using $D_{\rm par}$ is
comparable in cost to using $D_{\rm Wilson}$. 

\section{Acknowledgements}

This work has been supported in part by the Schweizerischer Nationalfonds.

\end{document}